\documentclass[doublecol]{epl2} 

\usepackage{graphicx}
\usepackage{dcolumn}
\usepackage{bm}

\title{Exchange-mediated dynamic screening in the integer quantum Hall regime}
\shorttitle{Exchange-mediated dynamic screening in the IQHE}

\author{Josef Oswald\inst{1} \and Rudolf A.\ R\"{o}mer\inst{2}}
\shortauthor{J. Oswald \etal}

\institute{%
 \inst{1} Institut f\"{u}r Physik, Montanuniversit\"{a}t Leoben, Franz-Josef-Strasse 18, 8700 Leoben, Austria\\
\inst{2} Department of Physics and Centre for Scientific Computing, University of Warwick, Coventry, CV4 7AL, UK 
}%

\date{$Revision: 1.5 $, compiled \today}


\abstract{
We study many-body interaction effects in the spatially-resolved filling factor ($\nu$) distribution for higher Landau levels (LLs) via self-consistent Hartree-Fock simulations in the integer quantum Hall (IQH) regime. Our results indicate a strong, interaction-induced tendency to avoid the simultaneous existence of partially filled spin-up and spin-down LLs. Rather, we find that such partially filled LLs consist of coexisting regions of full and empty LLs. At the boundaries between the regions of full and empty LLs, we observe edge stripes of nearly constant $\nu$ close to \emph{half filling}. This suggests that the exchange interaction induces a behavior similar to a Hund's rule for the occupation of the spin split LLs. The screening of the disorder and edge potential appears significantly reduced as compared to static Thomas-Fermi screening \cite{ChkSG92}. Our results are consistent with a local, lateral $\nu$ dependence of the exchange-enhanced spin splitting. Hence, on quantum-coherent length scales as probed here, the electron system of the IQH effect behaves similar to a non-interacting single particle system --- not because of the absence, but rather due to the dominance of many-body effects. 
}

\begin{document}
\maketitle


Since its discovery more than 35 years ago \cite {KliDP80}, the integer quantum Hall effect (IQHE) has attracted considerable research effort. The discovery of the fractional quantum Hall effect (FQHE), just a few years later \cite {TsuSG82}, did little to diminish these efforts and the IQHE still remains an area of ongoing investigations due to its inherent richness in quantum phenomena. While the FQHE is often discussed in terms of quasi particles (such as composite fermions \cite{Jai15}) in the highly correlated many-particle electron system \cite{Lau83}, it is widely believed that the physics of the IQHE is dominated by single-particle interaction \cite{WeiK11}. Up to date the widely accepted and state of the art model for screening in the IQHE regime has been developed by Chlovskii, Shklovskii and Glazman (CSG), who addressed the electrostatics of the edge channel region \cite{ChkSG92}. Their key ingredient is the different screening capability of partially and fully filled Landau Levels (LLs); instead of getting narrow quasi one-dimensional channels \cite{Hal82,But88,ChaC88} CSG find so-called compressible stripes (CS) of partially filled LLs that can become up to hundreds of nanometer wide. These stripes are separated by usually narrow, incompressible stripes of fully filled LLs. The main result of CSG is that almost the entire slope of the electrostatic edge potential is screened out and terraces in the uprising electrostatic edge potential are generated. At the same time the electron density changes continuously across the CS. 

The CSG approach has been very successful in explaining many of the intricate features of IQHE physics \cite{CobBF99,IlaMTS04,MarIVS04}.
However, recent scanning gate microscopy (SGM) experiments by Pascher et al.\ \cite{PasRIE14}, who investigated edge stripes passing quantum point contacts at high magnetic field, indicate that the screening behavior within the compressible stripes is much weaker than predicted by the CSG approach. Their results fit even more closely to a system of non-interacting single carriers. On the other hand, a recent investigation of the local nature of compressibility in the bulk of a IQHE sample strongly challenges existing single-particle theories \cite{KenSKO17}. Such results raise the question whether many-particle effects are indeed effectively absent in the IQHE regime, or, quite opposite, whether they are responsible for implying the physics of almost non-interacting single particles. 
The answer manifests itself, of course, in the screening properties of the electron system.  The CSG approach is based on classical electrostatics and many-body exchange effects are not included. It is possible to add an exchange-enhanced $g$-factor as a semi-empirical parameter \cite{KenSKD13}, which, however, might still not be sufficient to capture all relevant many-particle physics. We note that the existence of an exchange-enhanced $g$-factor is well known and also widely accepted \cite{Nomura2013a}.

In order to revisit the many-body screening effects of a two-dimensional electron system (2DES) in the IQHE, we utilize (a) a modeling of the many-particle groundstate, for which we use a fully self-consistent Hartree-Fock implementation discussed in Refs.\ \cite{SohOR09,SohR07,Soh07}; (b) we also need a model for electron transport close to equilibrium while driving the electron system out of equilibrium. In order to avoid any violation of the physics of coherent many-particle quantum transport, the model also must not make explicit use of single carrier flow. We have shown previously that our non-equilibrium network model \cite{OswO06} (NNM) is able to meet these requirements \cite{OswR15,Osw16}.
We find that partially filled LLs appear as a mixture of clusters of locally full and empty LLs in contradistinction to the Thomas-Fermi approximation by CSG. Stripes of nearly constant \emph{half-filling} exist at the boundaries of these clusters at higher LLs and serve as transmitting channels for charge transport.
Our results can be interpreted in terms of a laterally varying filling factor dependent enhanced $g$-factor. This reduces screening as compared to CSG in agreement with the aforementioned experimental observations \cite{PasRIE14}.


We consider a 2DES in the $(x,y)$-plane subject to a perpendicular magnetic field $\vec{B} = B\vec{e}_z$. Within the plane of the 2DES, we create a random, spatially correlated disorder potential using $N_{\rm I}$ Gaussian-type "impurities" to model a high-mobility heterostructure. The number of flux quanta piercing the 2DES of size $L\times L$ is given by $N_\phi=L^2/2\pi l_{\rm c}^2$, yielding a total number of $M = N_{\rm LL} N_\phi$ states per spin direction. The filling of the system is characterized by the filling factor $\nu = N_{\rm e}/N_\phi$, with $N_{\rm e}$ being the number of electrons in the system and areal density $n = N_{\rm e}/L^2$. The total Landau level density is given by $n_0 = eB/h$ and $l_{\rm c} = \sqrt{\hbar/eB}$ is the magnetic length.

\begin{figure}
\includegraphics[width=\columnwidth]{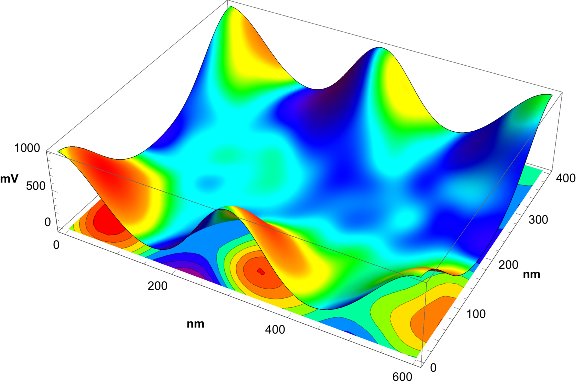}
\caption{\label{bare_pot} Bare potential for a Hall bar of size $600 \times 400$ nm$^2$. An \emph{edge confinement} potential is created by repulsive Gaussian peaks of $1$ V amplitude at the corners and in the middle of the longitudinal sides, leaving openings for current contacts at the ends and voltage probes along the sample. An additional potential \emph{disorder} of maximally $\pm 10$ mV is generated by randomly distributed Gaussians. This disorder is color coded onto the surface of the total potential with red representing $\sim -10$ mV and blue/purple representing $\sim 10$ mV. The contour plot in the plane also shows this disorder with the same color scheme and indicative contours.}
\end{figure}
Fig.~\ref{bare_pot} shows the bare model potential of a Hall bar structure of total size of $600\times 400$nm$^2$ that gets filled with $480$ electrons, such that $n = 2 \times 10^{11}$cm$^{-2}$. The solution of the Hartree-Fock model at temperature $T=1$ K for different $B$ fields delivers values of the local filling factors separately for electrons with spin-up and spin-down, i.e.\ $\nu_{\uparrow}$ and $\nu_{\downarrow}$, resp. Those are transferred to the NNM which calculates the lateral distribution of the experimentally injected non-equilibrium electrochemical potentials. No local quantities such as local conductivities or local Ohm’s laws are used up to this point, avoiding the (forbidden) possibility to track down the path of the individual carriers while moving from one current contact to the next. 

\begin{figure}
\includegraphics[width=\columnwidth]{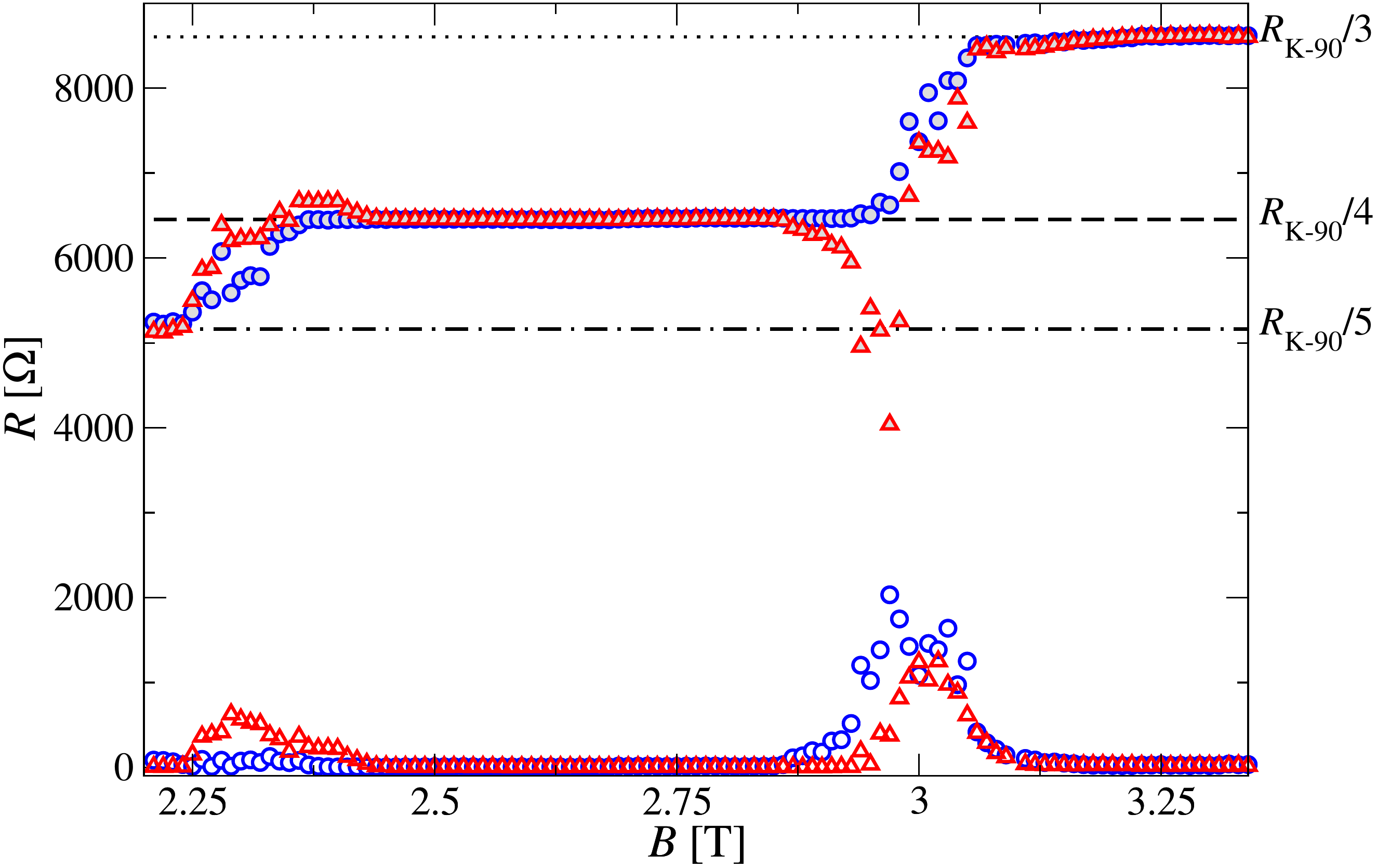}
\caption{\label{Rxxxy} Simulated $R_{xx}$ (lower traces) and Hall $R_{xy}$ (upper traces) as a function of magnetic field $B$ for the two longitudinal and Hall contact pairs plotted in different colors.}
\end{figure}
As can be seen from Fig.~\ref{Rxxxy} the obtained longitudinal, $R_{xx}$, and Hall, $R_{xy}$, resistances show the expected behavior,  exhibiting Hall plateaus while $R_{xx}=0$ and $R_{xx}$ peaks in the transition regime between QH plateaus. The average $n= 2 \times 10^{11}$cm$^{-2}$ corresponds to an effective density of about $2.5 \times 10^{11}$cm$^{-2}$ in the active region of the Hall sample structure and the peaks and plateau positions get shifted up in magnetic field. There appear strong resistance fluctuations in both $R_{xx}$ and $R_{xy}$ close to each QH transition as has to be expected for quantum Hall structures of mesoscopic size, as known also from experiments \cite{Simmons1991}.

\begin{figure*}
\includegraphics[width=0.31\textwidth]{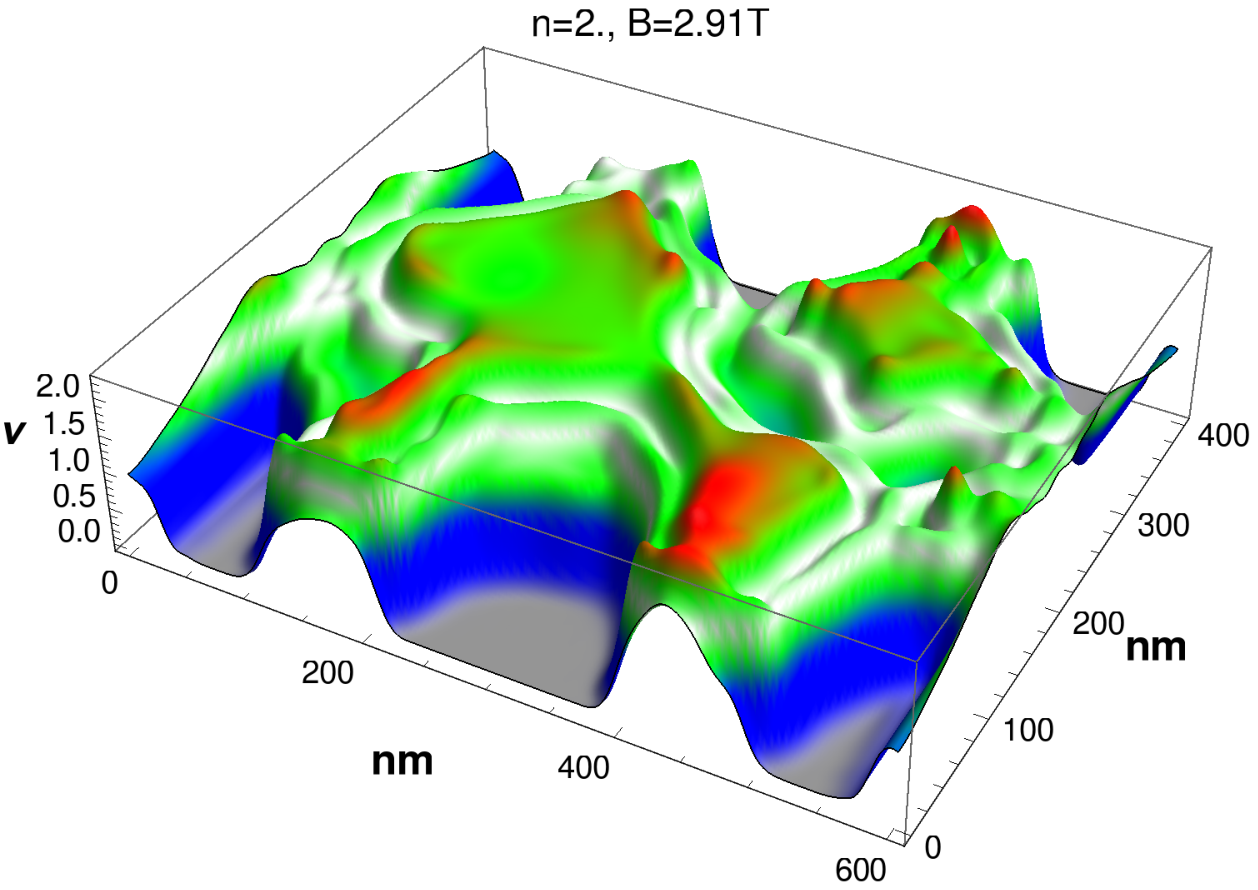}\hfill 
\includegraphics[width=0.31\textwidth]{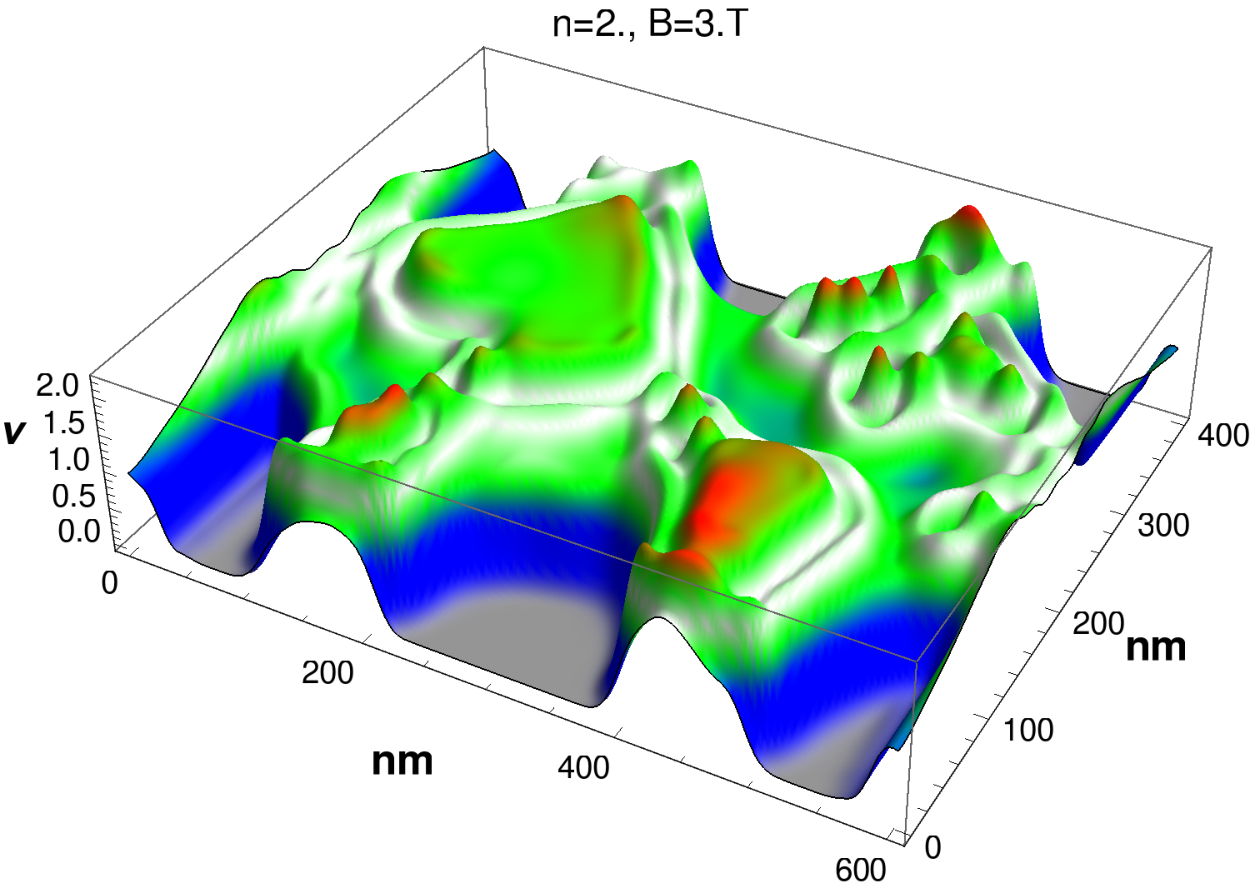}\hfill 
\includegraphics[width=0.31\textwidth]{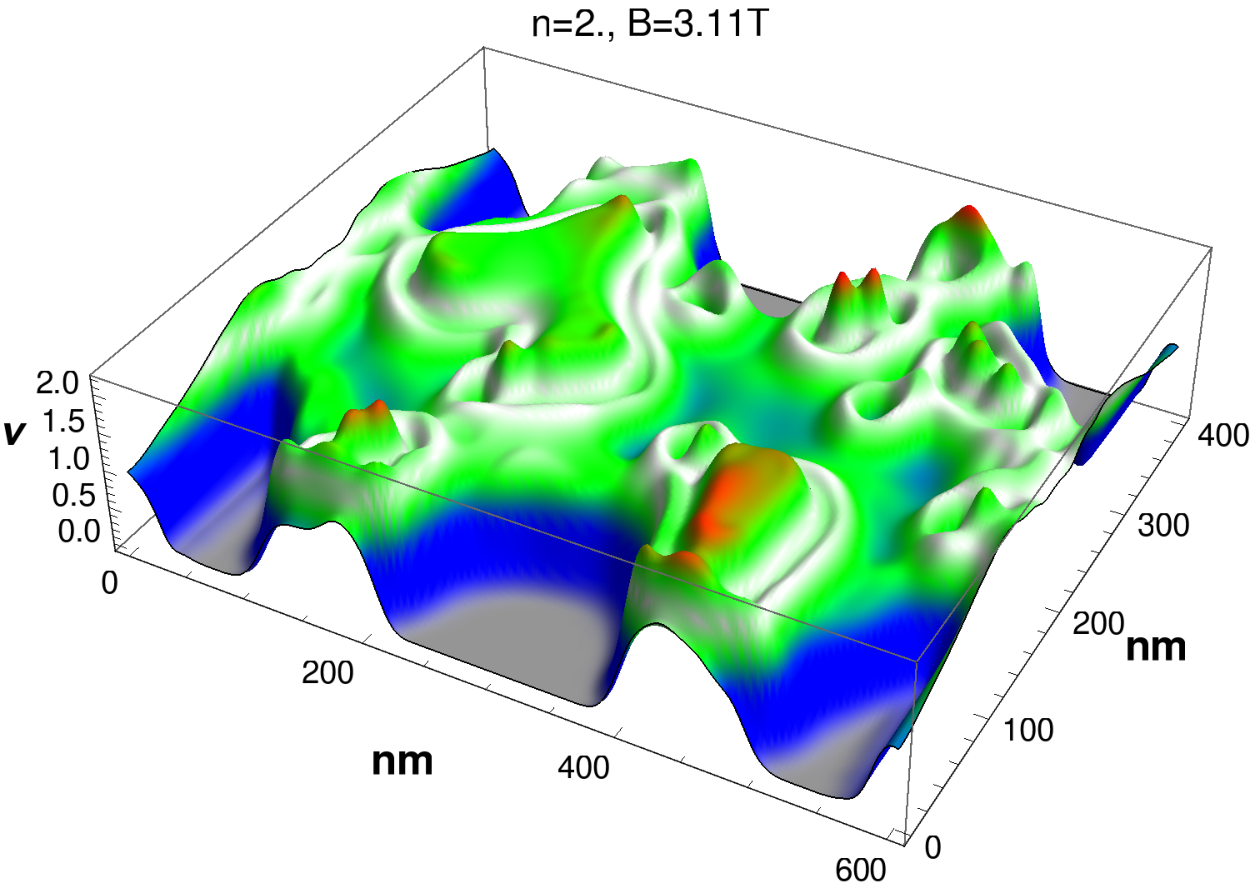}\\ 
\hspace*{-3ex}
(a)\includegraphics[width=0.29\textwidth]{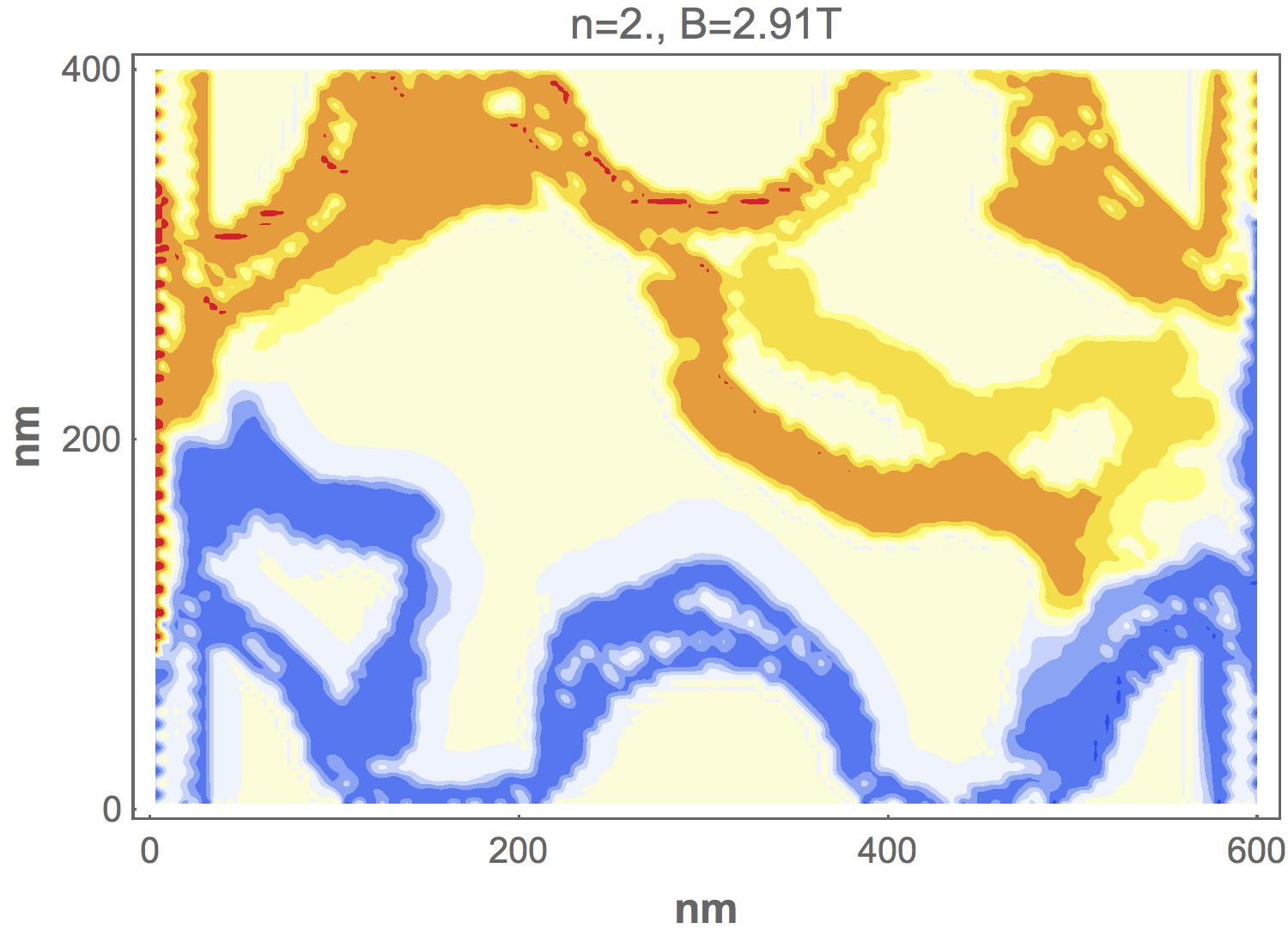}\hfill
(b)\includegraphics[width=0.29\textwidth]{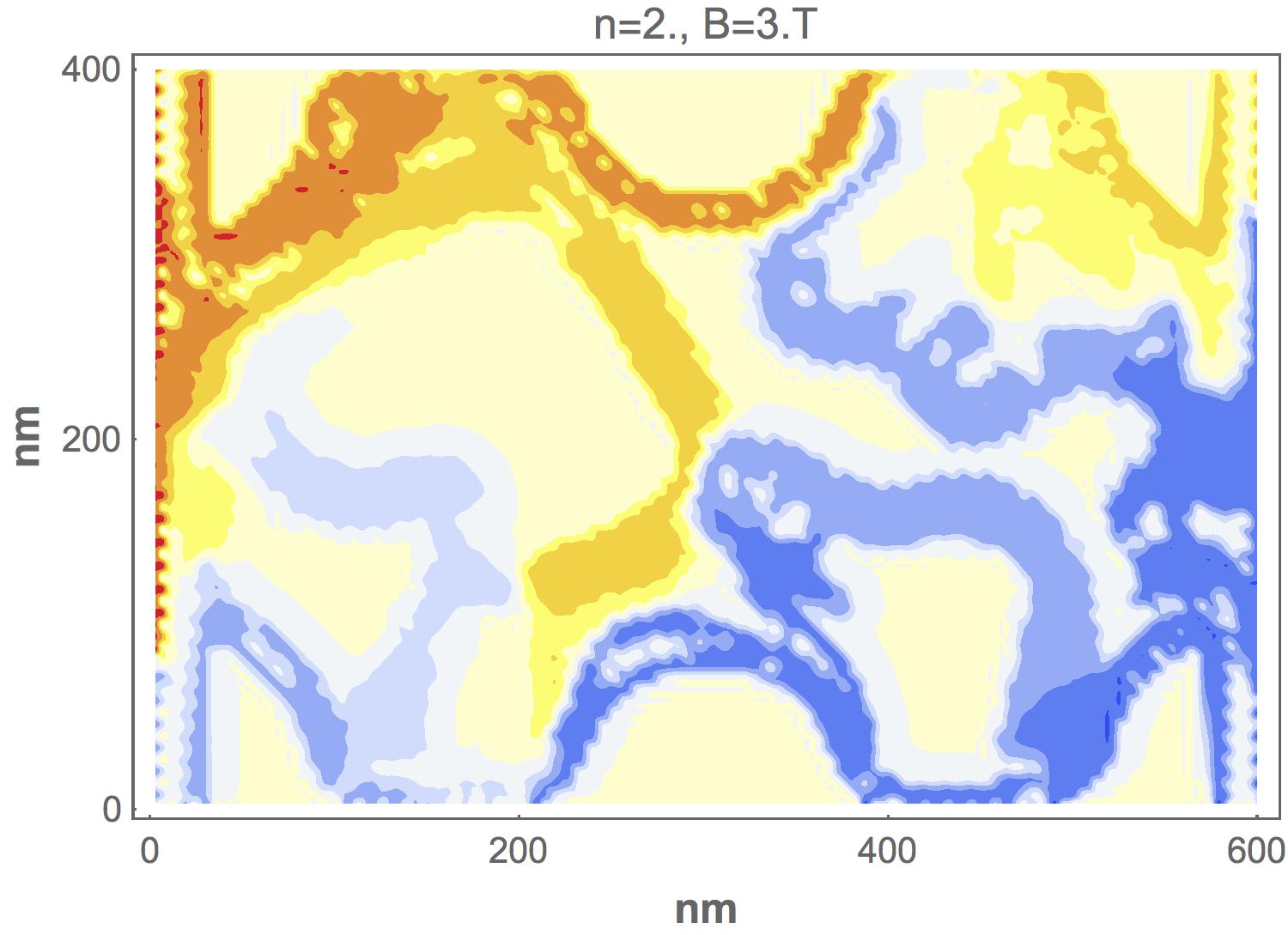}\hfill
(c)\includegraphics[width=0.29\textwidth]{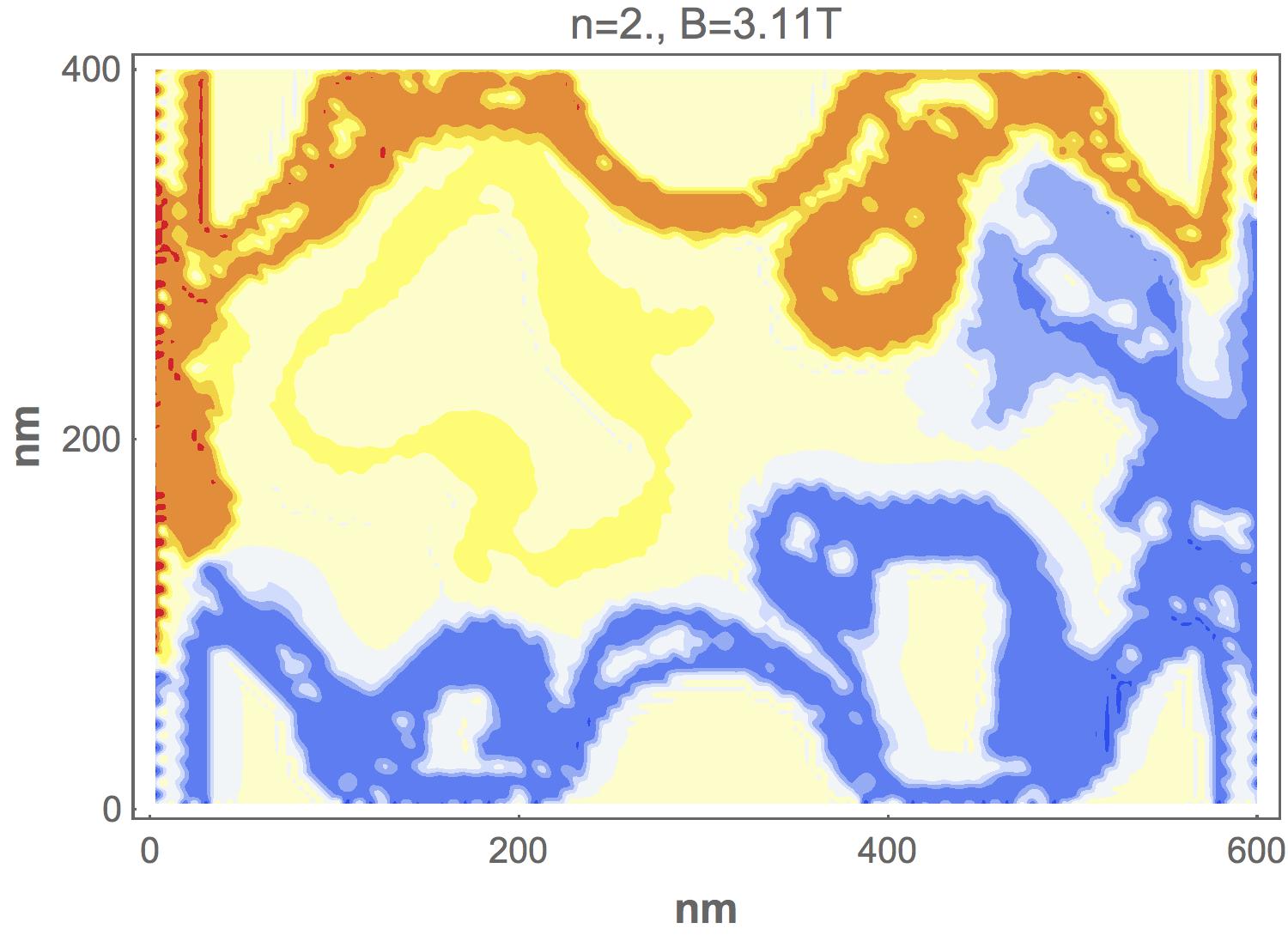}\\
\caption{\label{CDS1x3} (Top row) Lateral filling factor distribution of $\nu_{\uparrow}$ for the highest partly filled LL for the magnetic field sweep with $n=2\times 10^{11}$ cm$^{-2}$ shown in Fig.~\ref{Rxxxy}
(a) at the beginning of the  $\nu = 4 \rightarrow 3$ plateau transition at $B=2.91$ T, 
(b) in the middle of the plateau transition at $B=3$ T ($l_c \approx 15$ nm), and
(c) at the end of plateau transition at $B=3.11$ T. The colors represent the filling factor, where blue means the first LL for $\nu = 0 \rightarrow 1$, green the second LL for $\nu = 1 \rightarrow 2$ and red the third, all for spin-up electrons. The filling factor range close to $\nu = 1.5$  is highlighted in light gray in order to identify the stripes appearing close to the half filled top LL. The dark gray regions indicate the depleted edges regions with $\nu\approx 0$ corresponding to the large confining potentials in Fig.~\ref{bare_pot}.
\label{ECS2x3} (Bottom row) Lateral non-equilibrium distribution of the injected chemical potential $\mu$ at the corresponding three $B$ field strengths as in the top row. The colors represent $\mu$ in arbitrary units with high potential in dark orange and low potential in dark blue. 
\revision{(We note that channels enter the metallic voltage probes with an incoming potential and reappear with a possibly different potential according to mixing with back scattered bulk channels. The details of this mixing in the contacts are not visible in the plot, but handled correctly in the NNM.)}
}
\end{figure*}

Figure \ref{CDS1x3} (top row) displays the local filling factor of the top spin-up LL during the $\nu = 4 \rightarrow 3$ plateau transition. The spin-up level is higher in energy than the spin-down level and therefore depletion starts with the spin-up level that is initially at filling factor $\nu_{\uparrow} =2$, while for spin-down we have also $\nu_{\downarrow} =2$, giving in total a filling factor $\nu =4$. At the beginning of the plateau transition at $B= 2.91$ T (see also Fig.~\ref{Rxxxy}) the initially completely filled spin-up LL starts to break up in sub-regions of filling factor $\nu_{\uparrow} =2$ and regions of filling factor $\nu_{\uparrow} =1$, which gives an average filling factor between $\nu =4$ and $\nu =3$.  This goes on for the entire plateau transition and the clusters of $\nu_{\uparrow} =2$ shrink in size. At the boundaries of the $\nu_{\uparrow} =2$ clusters strips of filling factor close to $\nu_{\uparrow} =1.5$  are obtained in the spin-up LL.
In Fig.\ \ref{ECS2x3} (bottom row) we show the lateral distribution of the non-equilibrium electrochemical potential $\mu$ that is injected at the current contacts and obtained from the NNM for all three $B$ cases \revision{with clockwise chirality}. One can see that the potentials are transmitted along the half\revision{-odd} integer stripes appearing in Fig.\ \ref{CDS1x3} (a)--(c). If the potentials mix up at some locations in the bulk region this generates dissipation as it is the case in the transition regime of the IQHE \revision{ and indicated by the associated color changes in Fig.\ \ref{CDS1x3}}. While this happens only weakly in (a) for $B=2.91$ T and (c) for $B=3.11$ T, this mixing appears to be quite strong  for $B=3$ T in (b), which represents the QH transition close to the maximum of the $R_{xx}$-peak. 
\revision{We also observe dissipation at the voltage contacts. This is most clearly pronounced, e.g., in the upper right and the lower left voltage contacts in Fig.\ \ref{CDS1x3} (b). We note that the channels appear to merge into the metallic contact region and reappear with changed magnitude.}

The mechanism for the de-population of the spin-up LL upon increasing $B$ is now quite clear from Fig.~\ref{CDS1x3}. Instead of an overall uniform decrease of the carrier density, we find locally shrinking clusters of fully filled spin-up LL at $\nu_{\uparrow} =2$ and increasing regions of depleted spin-up LL at $\nu_{\uparrow} =1$, which on average give a continuously decreasing $\nu$. Therefore a total filling factor of, say, $\nu_{\uparrow} =1.5$ is made up by half of the area taken up by clusters of filling factor $\nu_{\uparrow} =2$ and half of the area taken up by regions of $\nu_{\uparrow} =1$. The same happens subsequently if the spin-down LL gets de-populated, which stays at $\nu_{\downarrow} =2$ while de-populating the top spin-up  LL. This is quite in contrast to the CSG model and is purely a result of many-particle interactions (and does not happen for a Hartree calculation, i.e.\ without the exchange term). But this suggests that even a half-filled LL may provide only poor screening as compared to a Thomas-Fermi-like behavior that would predict a continuous carrier density variation over all the sample area. In addition, at the boundaries between clusters of filling factor $\nu_{\uparrow} =2$ and $\nu_{\uparrow} =1$ there appear terraces of almost constant filling factor $\nu_{\uparrow} =1.5$, cp.\ Fig.~\ref{CDS1x3} (top row).  
\revision{The width of the half\revision{-odd} integer stripes appear to be of the order of the magnetic length $l_c$. We furthermore find two parallel stripes of width $l_c$ at the boundaries between clusters $\nu_{\uparrow} =3$ and $\nu_{\uparrow} =2$ (not shown here). This suggest to us that the origin of the stripes lies in the spatial dependence of the Landau states similar to what was measured for the behaviour of the local density of states \cite{HasCFS12}. The cluster boundaries and the boundaries of the fully filled LLs at the sample edge are the only regions where the many-particle electron system can exchange carriers close to equilibrium as in low excitation magneto transport experiments. As a consequence, these boundaries are experimentally observed as transport channels \cite{Osw16}.}


The observed behavior during population and de-population results from a tendency of the many-particle system to avoid the simultaneous appearance of partially filled spin-up and spin-down LLs. A similar behavior would result by applying Hund’s rule, which seems to be the driving force also for many-particle effects in the QHE regime. \revision{During filling, electrons continue to be added to the laterally growing region of the filled top spin-split LL, even if its boundary already extends into regions of classically not accessible elevated edge and random potentials. The process continues as long as possible and therefore the filling of the next upper level is considerably delayed at higher carrier densities as compared to single particle considerations.}

\begin{figure*}
(a)\includegraphics[angle=0,keepaspectratio=true,width=0.45\textwidth]{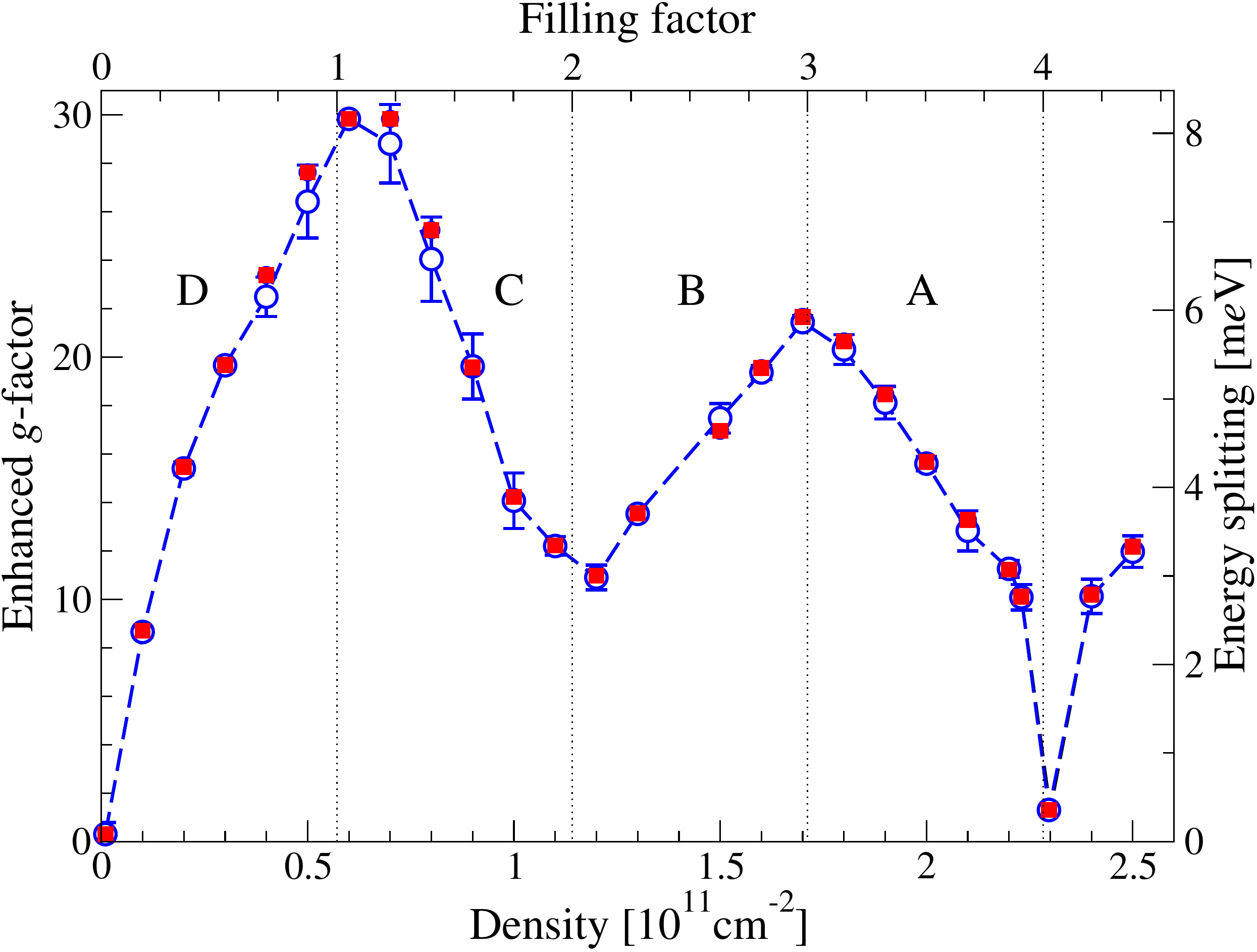}
(b)\includegraphics[angle=0,keepaspectratio=true,width=0.45\textwidth
]{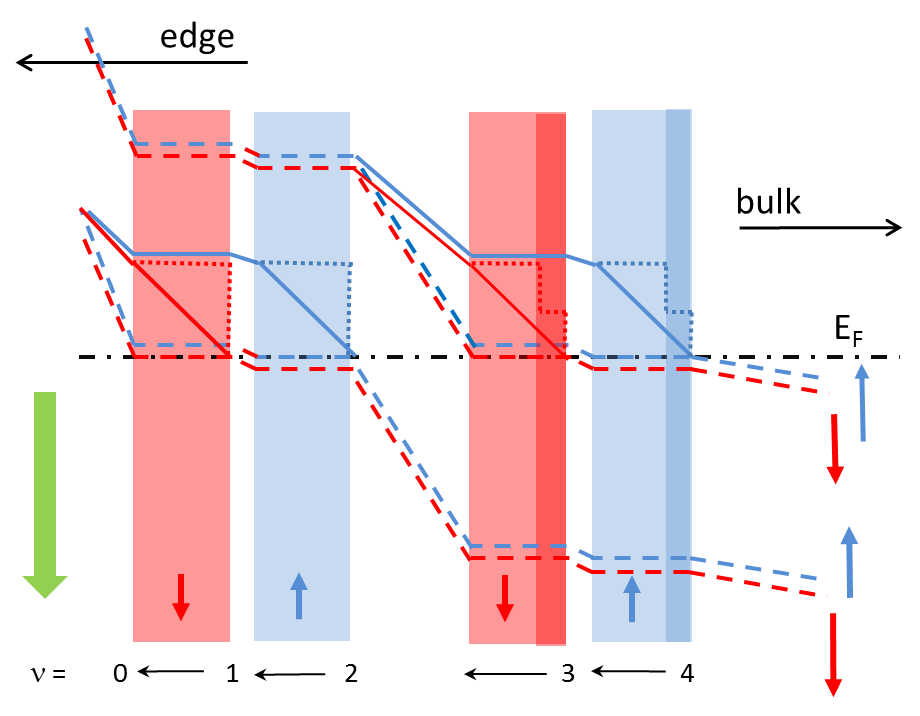}

\caption{\label{pushup}  
(a) Plot of the $g$-enhancement factor (alternatively $\Delta E$) as a function of $n$ (and $\nu$). The open (blue) circles denote the mean and error bars indicate the standard error while closed (red) squares show the median. The dashed line is a guide to the eye for the mean values. Labels A, B, C, and D are used in the discussion in the text to identify certain behaviours. 
(b) Schematic screened edge potential according to CSG including the bare spin splitting (dashed lines); the shaded regions mark the spin-down/spin-up compressible stripes, resp. The bold lines indicate the additional filling factor dependent exchange driven Zeeman energy if the CS according to CSG would remain stable. The  dotted lines indicate the additional Zeeman energy after carrier re-distribution. Red/blue indicate spin-down/up as specified by the arrows. The vertical green arrow indicates the direction of the $B$ field.}
\end{figure*}

Let us now discuss our findings using the language of an enhanced $g$-factor. In this way, we can make contact with the single-particle picture by including the majority of the many-body physics in the renormalized $g$. We have studied the energy splitting $\Delta E = g\ g_\mathrm{spin} \mu_B B$ where $\mu_{\rm B}$ denotes the Bohr magneton  $g_\mathrm{spin}\approx 2$ is the bare electron $g$-factor, and $g$ its enhancement factor, respectively. Averaging over $\Delta E$ between the occupied spin-down and the unoccupied spin-up states, we can compute $g$ as shown in Fig.\ \ref{pushup}(a). We see that there is indeed a considerable enhancement with $g \gg 1$ at odd $\nu$ while the enhancement drops for even $\nu$.
This enhancement now implies changes to the CSG picture as shown in Fig.\ \ref{pushup}(b). Starting from deep in the bulk, with a Fermi energy $\epsilon_F$ well above the spin-up LL2 (supposing an insulating bulk), we now see that upon going towards the edge via $\nu=4 \rightarrow 3$, while keeping the screened edge potential constant and continuously decreasing $n$ (as requested by CSG), the increased $g$-value implies a rise in the spin-up level energy shown as bold blue line (compare also branch A in Fig.\ \ref{pushup} (a)). This, of course, is not a stable situation because the up-rising spin-up level gets pushed above the Fermi level and initiates a self consistent carrier redistribution that leads to an almost abrupt drop of the filling factor to the next lower odd integer $\nu = 3$ while at the same time pushing up the spin-up level to the maximum g-enhancement at $\nu = 3$ (as shown schematically by the dotted blue line in the stripe $\nu=4 \rightarrow 3$). According to our interpretation of the self consistent HF data, this resembles an almost step like increase of the spin-up level at the boundary on the right side of the originally wide CS stripe, just leaving a narrow stripe of nearly half\revision{-odd} integer filling with g-enhancement according to $\nu = 3.5$ (marked by the narrow dark shaded blue bar).  
Conversely, for odd $\nu=3 \rightarrow 2$, the $g$-enhancement drops, and the two-spin split levels are getting closer and  the uprising spin-down level (bold red) accounts for this decrease of spin splitting (see also branch B in Fig.\ \ref{pushup}(a)). Again the resulting carrier re-distribution causes an almost abrupt jump to the next lower integer filling, which in this case is an even value at $\nu = 2$. Again a narrow half\revision{-odd} integer stripe in the step like uprising spin-down level is left (dotted red line and narrow dark shaded red bar). Now the g-enhancement almost vanishes until we enter the next CS at $\nu=2 \rightarrow 1$ where the whole discussion starts over again on the basis of branches C and D in Fig.\ \ref{pushup}(a). The only exception is the fact that such half\revision{-odd} integer features do not appear in the lowest LL. The same discussion as done above for the edge region, of course, also applies for the boundaries of any clusters in the bulk region, that are caused by the random potential fluctuation.

Results showing the behaviour of $\nu_{\uparrow}$ and $\nu_{\downarrow}$ upon varying $n$ can be found in the supplement \cite{QHsupplement} and reconfirm the above arguments. \revision{When} increasing the temperature we find that the fine features shown in Fig.\ \ref{CDS1x3} slowly vanish. Last, the features found at half\revision{-odd} integer filling are not present when we restrict our calculations to include only the Hartree term \cite{OswR17b}.

In conclusion, we show that the IQH regime is dominated by many-particle physics that acts towards re-establishing the narrow quasi one-dimensional channels of non-interacting single carriers as assumed by the early quite successful models of the IQHE \cite{Hal82,But88,ChaC88}. The most striking fact is that the many-body interactions replace the plateaus in the edge potential of CSG by narrow \emph{half\revision{-odd} integer features} in the local filling factor in higher LLs. 
The apparent modifications to the CSG model might have its reason in the fact, that for the reduced system size of our simulations, that so far is in the sub-micron range, the CSG model breaks down. The validity of CSG might hold only for structures that are significantly larger where a smooth average filling factor is a good representation for our observed mixture of clusters of full and empty LLs in the HF results. In this context our model possibly provides a so far missing link between the quite successful semiclassical CSG approach and the microscopic many-particle world. However, this aspect is the goal of the ongoing and future work. Considering the recent developments in locally-resolved studies in the QH regime \cite{HasSWI08,HasCFS12,PasRIE14} it seems likely that the half\revision{-odd} integer features can be detected experimentally in the near future.

\acknowledgments
We gratefully acknowledge funding from the Austrian Science Foundation FWF Project P19353-N16, EPSRC grant EP/J003476/1 and provision of computing resources through the MidPlus Regional HPC Centre Warwick UK (EP/K000128/1). UK data statement: Data is available via \cite{QHsupplement}.


\begin{thebibliography}{10}
\expandafter\ifx\csname url\endcsname\relax\def\url#1{\texttt{#1}}\fi

\bibitem{ChkSG92}
\Name{Chklovskii D.~B., Shklovskii B.~I. \and Glazman L.~I.} \REVIEW{Physical
  Review B}{46}{1992}{4026}.

\bibitem{KliDP80}
\Name{Klitzing K.~v., Dorda G. \and Pepper M.} \REVIEW{Physical Review
  Letters}{45}{1980}{494}.

\bibitem{TsuSG82}
\Name{Tsui D.~C., Stormer H.~L. \and Gossard A.~C.} \REVIEW{Physical Review
  Letters}{48}{1982}{1559}.

\bibitem{Jai15}
\Name{Jain J.~K.} \REVIEW{Annual Review of Condensed Matter
  Physics}{6}{2015}{39}.

\bibitem{Lau83}
\Name{Laughlin R.~B.} \REVIEW{Physical Review Letters}{50}{1983}{1395}.

\bibitem{WeiK11}
\Name{Weis J. \and von Klitzing K.} \REVIEW{Philosophical Transactions of the
  Royal Society of London A: Mathematical, Physical and Engineering
  Sciences}{369}{2011}{3954}.

\bibitem{Hal82}
\Name{Halperin B.~I.} \REVIEW{Physical Review B}{25}{1982}{2185}.

\bibitem{But88}
\Name{B{\"{u}}ttiker M.} \REVIEW{Physical Review B}{38}{1988}{9375}.

\bibitem{ChaC88}
\Name{Chalker J.~T. \and Coddington P.~D.} \REVIEW{Journal of Physics C: Solid
  State Physics}{21}{1988}{2665}.

\bibitem{CobBF99}
\Name{Cobden D.~H., Barnes C. H.~W. \and Ford C. J.~B.} \REVIEW{Physical Review
  Letters}{82}{1999}{4695}.

\bibitem{IlaMTS04}
\Name{Ilani S., Martin J., Teitelbaum E., Smet J.~H., Mahalu D., Umansky V.
  \and Yacoby A.} \REVIEW{Nature}{427}{2004}{328}.

\bibitem{MarIVS04}
\Name{Martin J., Ilani S., Verdene B., Smet J., Umansky V., Mahalu D., Schuh
  D., Abstreiter G. \and Yacoby A.} \REVIEW{Science}{305}{2004}{}.

\bibitem{PasRIE14}
\Name{Pascher N., R{\"{o}}ssler C., Ihn T., Ensslin K., Reichl C. \and
  Wegscheider W.} \REVIEW{Physical Review X}{4}{2014}{11014}.

\bibitem{KenSKO17}
\Name{Kendirlik E.~M., Sirt S., Kalkan S.~B., Ofek N., Umansky V. \and Siddiki
  A.} \REVIEW{Nature Communications}{8}{2017}{14082}.

\bibitem{KenSKD13}
\Name{Kendirlik E.~M., Sirt S., Kalkan S.~B., Dietsche W., Wegscheider W.,
  Ludwig S. \and Siddiki A.} \REVIEW{Scientific Reports}{3}{2013}{494}.

\bibitem{Nomura2013a}
\Name{Nomura S., Tamura H., Yamaguchi M., Akazaki T. \and Hirayama Y.}
  \REVIEW{Physical Review B}{87}{2013}{085318}.

\bibitem{SohOR09}
\Name{Sohrmann C., Oswald J. \and R{\"{o}}mer R.} \Book{{Quantum Percolation in
  the Quantum Hall Regime}} in \Book{Quantum and Semi-classical Percolation and
  Breakdown in Disordered Solids}, edited by \Name{Sen A.~K., Bardhan K.~K.
  \and Chakrabarti B.~K.} (Springer Berlin Heidelberg, Heidelberg) 2009 pp.
  1--31.

\bibitem{SohR07}
\Name{Sohrmann C. \and R{\"{o}}mer R.~A.} \REVIEW{New Journal of
  Physics}{9}{2007}{97}.

\bibitem{Soh07}
\Name{Sohrmann C.} \Book{{Interactions in the integer quantum Hall effect}}
  Ph.D. thesis University of Warwick (2007).

\bibitem{OswO06}
\Name{Oswald J. \and Oswald M.} \REVIEW{Journal of Physics: Condensed
  Matter}{18}{2006}{R101}.

\bibitem{OswR15}
\Name{Oswald J. \and R{\"{o}}mer R.~A.} \REVIEW{Physics
  Procedia}{75}{2015}{314}.

\bibitem{Osw16}
\Name{Oswald J.} \Book{{Linking Non-equilibrium Transport with the Many
  Particle Fermi Sea in the Quantum Hall Regime}} in \Book{Recent Advances in
  Quantum Dynamics}, edited by \Name{Bracken P.} (InTech, Rijeka) 2016 pp.
  131--163.

\bibitem{Simmons1991}
\Name{Simmons J.~A., Hwang S.~W., Tsui D.~C., Wei H.~P., Engel L.~W. \and
  Shayegan M.} \REVIEW{Physical Review B}{44}{1991}{12933}.

\bibitem{HasCFS12}
\Name{Hashimoto K., Champel T., Florens S., Sohrmann C., Wiebe J., Hirayama Y.,
  Roemer R.~A., Wiesendanger R. \and Morgenstern M.} \REVIEW{Physical Review
  Letters}{109}{2012}{116805}.

\bibitem{QHsupplement}
\Name{R{\"{o}}mer R.~A. \and Oswald J.}
  \Book{{http://wrap.warwick.ac.uk/id/eprint/86004}} (2017).

\bibitem{OswR17b}
\Name{Oswald J. \and R{\"{o}}mer R.~A.} \Book{{Interaction-driven enhanced
  g-factor in the IQHE}} (2017).

\bibitem{HasSWI08}
\Name{Hashimoto K., Sohrmann C., Wiebe J., Inaoka T., Meier F., Hirayama Y.,
  R{\"{o}}mer R.~A., Wiesendanger R. \and Morgenstern M.} \REVIEW{Physical
  Review Letters}{101}{2008}{256802}.

\end{thebibliography}
\end{document}